\newcommand{\vecR}{{\bf R}}
\newcommand{\vecL}{{\bf L}}
\newcommand{\vecM}{{\bf M}}
\newcommand{\vecr}{{\bf r}}
\newcommand{\vecv}{{\bf v}}
\newcommand{\vecp}{{\bf p}}

\newcommand{\vecnl}{{\bf 0}}

\newcommand{\unvec}[1]{\mbox{{\bf e}$_{#1}$}}

\newcommand{\vecomega}{\mbox{\boldmath$\omega$}}

\documentclass[12pt]{article}

\title{Average Angular Velocity}
\author{Hanno
Ess\'en \\Department of Mechanics \\Royal Institute of
Technology  \\ S-100 44 Stockholm, Sweden}  \date{1992,
December}
\begin{document}
\maketitle
\begin{abstract}
\normalsize This paper addresses the problem of the separation of
rotational and internal motion.  It introduces the concept of
average angular velocity as the moment of inertia weighted average
of particle angular velocities. It extends and elucidates the
concept of Jellinek and Li (1989) of separation of the energy of
overall rotation in an arbitrary (non-linear) $N$-particle system.
It generalizes the so called Koenig's theorem on the two parts of
the kinetic energy (center of mass plus internal) to three parts:
center of mass, rotational, plus the remaining internal energy
relative to an optimally translating and rotating frame.
 \vskip 1cm \noindent Published in: {\em
European Journal of Physics} {\bf 14}, pp.201-205, (1993).
\end{abstract}

\newpage
\section{Introduction}
The motion of a rigid body is completely characterized by its
(center of mass) translational velocity and its angular velocity
which describes the rotational motion.
Rotational motion as a phenomenon is, however, not restricted
to rigid bodies and it is then a kinematic problem to define
exactly what the rotational motion of the system is. This paper
introduces the new concept of `average angular velocity' as the
solution this problem and discusses some applications briefly.

The average angular velocity concept is closely analogous to
the concept of center of mass velocity. For a system of
particles the center of mass velocity is simply the mass
weighted average of the particle velocities. In a similar way
the average angular velocity is the moment of inertia weighted
average of the angular velocities of the particle position
vectors.

The separation of rotational motion from the internal degrees of
freedom of a system is of interest in a wide variety of
applications. Among the more obvious are vibration-rotation
coupling in polyatomic molecules and the understanding of
biomechanical phenomena involving falling cats, figure skaters,
springboard divers etc. The basic theoretical work on the subject
is Carl Eckart's (1935) whose work has been extended and
elucidated by many authors, for example Sayvetz (1939)
and Ess\'en (1979). Biomechanical aspects have been
discussed  by Frohlich (1979) from a
multibody dynamic point of view, and by Ess\'en (1981)
from the Eckart point of view.

Considering the maturity of the subject one might think that
all basic theoretical results are quite old, but a letter on the
subject was published as late as 1989 by Jellinek and Li
(1989). They showed that one can define an angular
velocity vector and separate out a rotational part of the
energy, for an arbitrary (nonlinear) system of particles,
without the use of the  rigid reference configuration that is
needed in the Eckart construction.

In this paper I present some new developments based on ideas
related to those of Jellinek and Li. In particular I introduce
the concept of {\em average angular velocity\/} as the weighted
average of particle angular velocities with moments of inertia
as weights. Especially for the elementary, but pedagogically
important, case of fixed axis rotation this leads to simple and
conceptually powerful results. The relevance of these results
to spinning athletes and turning cats are briefly indicated.

Later sections of the paper treat the three dimensional case.
It is shown that the separation of a rotational part of the
kinetic energy can be done in a way closely analogous to the
well known split of the kinetic energy to an overall center
of mass translational part plus the kinetic energy of
internal motion relative to the center of mass system.
A second split is thus done, now to an overall rotational
energy plus a remaining part corresponding to motions in a
rotating center of mass system, the rotation of which is given
by the average angular velocity.

Throughout this paper I consider a system of $N$ particles with
masses $m_k$ and position vectors $\vecr_k$ ($k=1,\ldots,N$).
A distance from the $Z$-axis is denoted by $\rho$, and a
distance from the origin by $r$, the azimuthal angle is
$\varphi$ and the angle to the positive $Z$-axis is $\theta$.

\section{Average Angular Velocity Around a Fixed Axis}
 The $z$-component
 of the angular  momentum, $L_z$, is by definition
\begin{equation}
L_{z}=\sum_{k=1}^{N}m_k(\vecr_k\times\vecv_k)
\cdot\unvec{z}
 =\sum_{k=1}^{N} m_k (x_k \dot y_k - y_k \dot x_k).
\end{equation}
 Here the $Z$-axis has a fixed direction and is either
fixed in an inertial system or fixed in the center of mass of the
system. We now introduce  cylindrical  (polar) coordinates
$\rho$, $\varphi$, and $z$. In terms of these we have for the
position vectors $\vecr_k = \rho_k\,{\bf e}_{\rho_k}+ z_k\,
\unvec{z}$ and for the velocities $\dot\vecr_k =
\dot\rho_k\,{\bf e}_{\rho_k}+ \rho_k
\dot\varphi_k\,{\bf e}_{\varphi_k}+ \dot z_k\,\unvec{z}$. This
gives us
\begin{equation} (\vecr_k \times
\vecv_k)\cdot\unvec{z} = [(\rho_k\,{\bf e}_{\rho_k}+
z_k\,\unvec{z})\times( \dot\rho_k\,{\bf e}_{\rho_k}+ \rho_k
\dot\varphi_k\,{\bf e}_{\varphi_k} +  \dot
z_k\,\unvec{z})]\cdot\unvec{z}
 = \rho_k^2\dot\varphi_k,
\end{equation}
so that
\begin{equation}
L_{z}(t) = \sum_{k=1}^{N} m_k \rho_k^2(t)\,\dot\varphi_k(t).
\end{equation}
If we now define the average angular velocity of the
system, around the $Z$-axis, by
\begin{equation}
\label{eq.average.angular.velocity}
\omega_{\rm av}(t) \equiv \langle \dot\varphi(t) \rangle_t =
\frac{ \sum_{k=1}^{N}  m_k \rho_k^2 (t)\, \dot\varphi_k(t) }{
\sum_{k=1}^{N}  m_k \rho_k^2 (t)}
\end{equation}
and the (instantaneous) moment of inertia, with respect to the
$Z$-axis, by
\begin{equation}
\label{eq.moment.of.inertia.w.r.to.z.axis}
J_z \equiv \sum_{k=1}^{N}  m_k \rho_k^2
\end{equation}
we see that we get
\begin{equation}
\label{eq.ang.momentum.w.r.z.axis.as.prod.of.Jz.and.phidot.av}
L_{z} = J_z \omega_{\rm av}.
\end{equation}
If all particles have the same  angular  velocity,
$\dot\varphi$, then, of course, $\omega_{\rm av} = \dot\varphi$.
This happens, in particular, if  the system of particles is
rigid and rotates around the $Z$-axis, but also more generally
for any type of motion that obeys $\dot\varphi_k = \dot\varphi$
with arbitrary $\dot\rho_k$ and $\dot z_k$. For these cases one
gets the standard result that $L_{z} = J_z \dot\varphi$.

We can now apply these results to the $z$-component of the
angular momentum principle ($\dot\vecL=\vecM$) in the form
\begin{equation}
\label{eq.ang.momentum.principle.z.comp.w.r.to.origin}
\dot L_{z} = M_{z} .
\end{equation}
The general result is that
\begin{equation}
\label{eq.z.ang.momentum.principle.with.L.eq.Jphidot.av}
\dot J_z \omega_{\rm av}  + J_z \dot\omega_{\rm av}
= M_{z} .
\end{equation}
If the angular velocity is  well defined we can replace
$\omega_{\rm av}$ by $\dot\varphi$ in this expression. If we
furthermore assume that the body is rigid so that
$J_z=\,$constant,  equation
(\ref{eq.z.ang.momentum.principle.with.L.eq.Jphidot.av}) reduces
to the standard result $J_z \ddot\varphi = M_{z}$.

If there is no external  moment (or `torque') with  respect to
the $Z$-axis, so that $M_{z}=0$, then the $z$-component of the
angular momentum vector will be conserved, $L_{z}=\,$constant,
and, in view of equation
(\ref{eq.ang.momentum.w.r.z.axis.as.prod.of.Jz.and.phidot.av}),
we find that
\begin{equation}
\label{eq.z.ang.momentum.conserved}
J_z(t)\, \omega_{\rm av}(t)=\mbox{constant}.
\end{equation}
This then says that a large moment of
inertia implies small average angular velocity and vice versa.
Here it is not assumed that there is rigidity or
even a definite angular velocity. It is this more general
form of the standard text book formula, $J_z \dot\varphi
=\,$constant, that is actually `used' by  springboard divers and
figure skaters.

\section{The Cat Landing on its Feet}
For some arbitrary quantity $\gamma_k$ the averaging of
equation (\ref{eq.average.angular.velocity}) can be written
\begin{equation}
\langle \gamma(t) \rangle_t \equiv
\frac{ \sum_{k=1}^{N}  m_k \rho_k^2(t)\, \gamma_k(t) }{
\sum_{k=1}^{N}  m_k \rho_k^2(t)} .
\end{equation}
A question of interest is to what extent the average angular
velocity can be understood as the time derivative of the
`average angle', $\langle \varphi \rangle_t$, of the system.
The subscript $t$ on the averaging bracket is meant as a
reminder that the weights in the averaging, $m_k \rho_k^2(t)$,
are time dependent and this means that the time derivative of
an average will not be the same as the average of a time
derivative.

If we take the time derivative of
\begin{equation}
\varphi_{\rm av} \equiv \langle \varphi \rangle_t
\end{equation}
a simple calculation shows that
\begin{equation}
\omega_{\rm av}  =\frac{d \varphi_{\rm av}}{dt} + 2\langle
\frac{ \dot\rho}{\rho} (\varphi_{\rm av} - \varphi ) \rangle_t.
\end{equation}
The average angular velocity is thus {\em not} simply the time
derivative of the average angle. This is, of course, essential
if a cat, dropped upside-down, is to be able to land on its
feet. Equation (\ref{eq.z.ang.momentum.conserved}) shows that
if $\omega_{\rm av}=0$ initially, it will remain zero in the
absence of external torque. The above equation reassures one
that the cat, nevertheless, can change its average angle by a
proper combination of angular and radial motions.

\section{The Average Angle Concept}
The concept of an `average angle' requires some comment. The
value of this angle will, of course, depend on the direction of
the fixed reference direction (the $X$-axis). It will also
depend on whether one thinks of $\varphi$ as going from $-\pi$ to
$\pi$ or if it goes from $0$ to $2\pi$, when one assigns
initial values to the $\varphi_k$. That is, $\varphi_{\rm av}$
depends on whether the necessary $2\pi$ jump comes at the
negative or at the positive $X$-axis, respectively. For a
cylinder the initial average angle will be zero in the former
case and $\pi$ in the latter. The actual value of the average
angle therefore has little physical meaning; its significance
comes from the fact that it defines a reference direction in
the particle system. Then, when the system has moved, it will
tell how large, on average, the net turn has been.

The time dependence of the averaging naturally vanishes if the
radii, $\rho_k$, are constant. It is interesting to note that it
also vanishes in the more general case when the time
dependencies of the (cylindrical) radii are of the form
\begin{equation}
\rho_k(t) = f(t)\, d_k,
\end{equation}
where $f(t)$ is some (positive) function of time and the $d_k$
are constants. The average then becomes
\begin{equation}
\langle \gamma(t) \rangle_t =
\frac{ \sum_{k=1}^{N}  m_k  f^2(t)\, d_k^2\,\gamma_k(t) }{
\sum_{k=1}^{N}  m_k  f^2(t)\,d_k^2} =
\frac{ \sum_{k=1}^{N}  m_k d_k^2\, \gamma_k(t) }{
\sum_{k=1}^{N}  m_k d_k^2 } = \langle \gamma(t) \rangle .
\end{equation}
For this case then, when the cylindrical radial motion is a
common `scaling', the time derivative operator commutes with the
operation of taking the average. Consequently the average
angular velocity will be the time derivative of the  average
angle and similarly for angular acceleration.

\section{K\"onig's Theorem}
The kinetic energy of an $N$-particle system is given by the sum
\begin{equation}
T = \frac{1}{2} \sum_{k=1}^{N} m_k \dot{\bf r}_k \cdot
\dot{\bf r}_k .
\end{equation}
If one introduces the center of mass position vector
\begin{equation}
\vecR \equiv \frac{ \sum_{k=1}^{N}  m_k \vecr_k }{
\sum_{k=1}^{N}  m_k } ,
\end{equation}
and then re-writes the position vectors of the particles as
\begin{equation}
\vecr_k = \vecR + \vecr'_k,
\end{equation}
the kinetic energy is seen to fall into two parts.
One part corresponds to the motion of the center of mass of
the system while the remaining part is due to the motion of
the particles relative to the center of mass system:
\begin{equation}
T = \frac{1}{2}m \dot\vecR \cdot\dot\vecR \;+\; T'.
\end{equation}
Here $m=\sum_{k=1}^{N} m_k$ is the total mass and $T'$, which is
given by
\begin{equation}
T' = \frac{1}{2} \sum_{k=1}^{N} m_k \dot{\bf r}'_k \cdot
\dot{\bf r}'_k,
\end{equation}
vanishes if the particles do not move relative to a
reference frame in which the center of mass is at rest.
The absence of cross terms is due to the fact that the sums,
$\sum_{k=1}^{N} m_k \vecr'_k = \sum_{k=1}^{N} m_k \dot\vecr'_k
= \vecnl$,
over the center of mass system position vectors and velocities,
vanish. This result is sometimes called K\"onig's theorem
(see Synge and Griffith 1970), or the `law of the two parts of
the kinetic energy', if it is given any name at all. It is this
result that can be taken a step further according to Jellinek
and Li (1989) in the sense that $T'$ can be split into
two parts, one corresponding to an overall rotation and the rest
corresponding to the motion relative to the rotating system.
This is shown in the next section.

\section{The Average Angular Velocity Vector}
The quantities of this section may be thought of as referring
to the center of mass system but we will drop the primes of the
previous section.  Introduce spherical coordinates $(r_k,
\theta_k, \varphi_k)$ for particle $k$ and corresponding moving
basis vectors $({\bf e}_{r_k}, {\bf e}_{\theta_k},  {\bf
e}_{\varphi_k})$. The position vector, ${\bf r}_k$, of particle
$k$ is then
\begin{equation}
{\bf r}_k = r_k \, {\bf e}_{r_k}.
\end{equation}
and the velocity of the particle is
\begin{equation}
\dot{\bf r}_k = \dot r_k \, {\bf e}_{r_k} +
 {\mbox{\boldmath $\omega$}}_k \times {\bf r}_k.
\end{equation}
Here ${\mbox{\boldmath $\omega$}}_k$ is the angular
velocity of the position vector of particle $k$. It is
given by
\begin{equation}
 {\mbox{\boldmath $\omega$}}_k =
\dot\varphi_k \cos\theta_k\,{\bf e}_{r_k}- \dot\varphi_k
\sin\theta_k\, {\bf e}_{\theta_k} +\dot\theta_k\,
{\bf e}_{\varphi_k} = \dot\varphi_k\,{\bf
e}_z + \dot\theta_k\, {\bf e}_{\varphi_k} .
\end{equation}
 The kinetic energy of the $N$-particle system is now
\begin{equation}
 T' =  \frac{1}{2} \sum_{k=1}^{N} m_k
(\dot r_k \, {\bf e}_{r_k} +
 \vecomega_k \times {\bf r}_k) \cdot
(\dot r_k \, {\bf e}_{r_k} +
 \vecomega_k \times {\bf r}_k)
\end{equation}
\begin{equation}
 =\frac{1}{2} \sum_{k=1}^{N} m_k \dot r_k^2
+ \frac{1}{2} \sum_{k=1}^{N} m_k (
\vecomega_k \times {\bf r}_k) \cdot(
\vecomega_k \times {\bf r}_k)
\end{equation}
\begin{equation}
\label{eq.kinet.energ.rad.plus.part.rot}
 =\frac{1}{2} \sum_{k=1}^{N} m_k \dot r_k^2 + \frac{1}{2}
\sum_{k=1}^{N}  \hat J_k \vecomega_k  \cdot \vecomega_k
\end{equation}
where $\hat J_k$ is the contribution of particle $k$ to
the (instantaneous) inertia tensor $\hat J$ of the
system. The matrix components of the inertia tensor
$\hat J_k$ in the basis  $({\bf e}_{r_k}, {\bf
e}_{\theta_k}, {\bf e}_{\varphi_k})$ are given by
\begin{equation}
\hat J_k =
\left(
\begin{array}{ccc}
0 & 0 & 0 \\
0 & m_k r_k^2 & 0 \\
0 & 0 & m_kr_k^2
\end{array} \right)
\end{equation}
Using this one easily verifies that the sum of the $k$th
terms of the sums of formula
(\ref{eq.kinet.energ.rad.plus.part.rot})
 gives the usual expression for the kinetic
energy of particle $k$ in spherical coordinates: $T_k =
\frac{1}{2} m_k [\dot r_k^2 + r_k^2(  \dot\varphi_k^2
\sin\theta_k +\dot\theta_k^2  ) ]$.

Below we will need to manipulate sums of terms like those
in the above expression for the kinetic energy. One
must then remember that the position dependent basis
vectors  $({\bf e}_{r_k}, {\bf
e}_{\theta_k}, {\bf e}_{\varphi_k})$ are  different for each
particle (as indicated by the index $k$).
In order to proceed we therefore return to a common Cartesian
basis in the expression $\frac{1}{2} \sum_{k=1}^{N}
\hat J_k{\mbox{\boldmath $\omega$}}_k  \cdot
{\mbox{\boldmath $\omega$}}_k $. The  inertia tensor $\hat
J_k$ will then have the matrix components
\begin{equation}
\hat J_k = m_k \left[ \left(
\begin{array}{ccc}
r_k^2 & 0 & 0 \\
0 & r_k^2 & 0 \\
0 & 0 & r_k^2
\end{array} \right) - \left(
\begin{array}{ccc}
  x_k  x_k  &   x_k  y_k  &   x_k  z_k  \\
  y_k  x_k  &   y_k  y_k  &   y_k  z_k  \\
  z_k  x_k  &   z_k  y_k  &   z_k  z_k
\end{array} \right) \right]
\end{equation}
and the sum of these
gives
\begin{equation}
\hat J =\sum_{k=1}^{N} \hat J_k= \sum_{k=1}^{N} m_k \left(
\begin{array}{ccc}
 y_k^2 +z_k^2  & -x_k y_k & -x_k z_k \\
-y_k x_k &  x_k^2 + z_k^2  & -y_k z_k \\
-z_k x_k &  -z_k y_k  & x_k^2 + y_k^2
\end{array} \right)
\end{equation}
i.e.\ the usual (instantaneous) inertia tensor for the
system of particles.

We now define the average angular velocity vector,
$\vecomega_{\rm av}$, by
\begin{equation}
\label{eq.aver.ang.vel.vect}
\hat J \vecomega_{\rm av} \equiv \sum_{k=1}^{N} \hat J_k
\vecomega_k,
\end{equation}
as the inertia tensor weighted average of the individual
particle angular velocity vectors. Here it is necessary that
$\hat J$ is invertible so that one can solve for $\vecomega_{\rm
av}$ by multiplying to the left by $\hat J^{-1}$. This means
that the particles of the system may not lie on a line since
then the inertia tensor is singular.

 If we now denote by $\vecomega'_k$ the angular velocity vector
of particle $k$ relative to the reference system that rotates
with the average angular velocity we have
\begin{equation}
\vecomega_k =\vecomega_{\rm av} + \vecomega'_k,
\end{equation}
since angular velocity vectors are additive. These relative
angular velocities  fulfill
\begin{equation}
\sum_{k=1}^{N} \hat J_k \vecomega'_k = \vecnl,
\end{equation}
so a calculation completely analogous to that leading to
K\"onig's theorem gives
\begin{equation}
T' = \frac{1}{2}  \hat J \vecomega_{\rm av}
  \cdot \vecomega_{\rm av}\; +\; T''.
\end{equation}
Here
\begin{equation}
T'' =\frac{1}{2} \sum_{k=1}^{N} m_k \dot r_k^2 + \frac{1}{2}
\sum_{k=1}^{N}  \hat J_k \vecomega'_k  \cdot \vecomega'_k
=  \frac{1}{2} \sum_{k=1}^{N} m_k {\bf v}''_k \cdot
{\bf v}''_k
\end{equation}
is the kinetic energy relative to a reference frame that rotates
with the average angular velocity (around the fixed center of
mass) and $\vecv''_k$ is the velocity of particle $k$ as
measured in this frame.

\section{Remarks on the Conservation Laws}
One notes that formula
(\ref{eq.aver.ang.vel.vect}) is nothing but the total angular
momentum, \vecL, of the system (with respect to the center of
mass):
\begin{equation}
 \hat J \vecomega_{\rm av} = \sum_{k=1}^{N} \hat J_k
\vecomega_k = \vecL.
\end{equation}
It is thus analogous to the formula
\begin{equation}
m \dot\vecR = \sum_{k=1}^{N} m_k \dot\vecr_k = \vecp
\end{equation}
for the total linear momentum, \vecp, of the system. Using the
linear and angular momenta the total kinetic energy of a system
can now be written
\begin{equation}
T = \frac{1}{2m} \vecp \cdot \vecp\; +\; \frac{1}{2} \hat J^{-1}
\vecL \cdot \vecL \; + \; T''.
\end{equation}
Here $m$ is the total mass, which is always constant, and \vecp\
the total linear momentum, which is constant in the absence of
external force. Thus in the absence of a net external force on
the system the first term in this expression for $T$ is
constant. What about the second term? In the absence of a net
external moment (of force)  on the system \vecL\ is constant,
but the (inverse) inertia tensor $\hat J^{-1}$ is, in
general, not since it depends on the `shape' of the system.

In an isolated body, such as a star or planet, one can expect
that internal dissipative forces, in the long run, will make
the internal relative motions zero so that $T'' \rightarrow 0$.
Such bodies will thus end up having only center of mass
translational and average rotational energy and both will then
be constant.

\section{Conclusions}
The fixed axis formalism at the beginning of this
paper is simple and useful enough to be included in even fairly
elementary texts, though I believe it is presented here for the
first time in the international literature. I have introduced
part of it in course notes that are used in the school of
engineering physics at KTH.  The point of view that the
Jellinek and Li (1989) separation of a rotational part
of the kinetic energy is analogous to the well known separation
of center of mass translational energy, is pedagogically useful
at higher levels, and the general ideas deserve to be better
known. It is the opinion of the author that most of the
equations and results of this paper, in fact, belong in a
comprehensive advanced mechanics course.

\newpage
\noindent
{\Large\bf Acknowledgments}\\
This work has been supported by the
Swedish Research Council for Engineering Sciences (TFR) and
the G\"oran Gustafsson Foundation.

\end{document}